\documentstyle[a4,12pt]{article}
\begin{document}
\parskip=5pt plus 1pt minus 1pt

\begin{flushright}
\bf DPNU-96-39 \\
August 1996
\end{flushright}

\vspace{0.2cm}

\begin{center}
{\large\bf Implications of the Quark Mass Hierarchy on Flavor Mixings} 
\end{center}

\vspace{0.3cm}

\begin{center}
{\sc Zhi-zhong Xing} \footnote{Electronic address: xing@eken.phys.nagoya-u.ac.jp}
\end{center}
\begin{center}
{\it Department of Physics, Nagoya University, Chikusa-ku, Nagoya 464-01, Japan}
\end{center}

\vspace{1.3cm}

\begin{abstract}
We stress that the observed pattern of flavor mixings can be 
partly interpreted by the quark mass hierarchy without the assumption
of specific quark mass matrices. The quantitatively proper relations between 
the Kobayashi-Maskawa matrix elements and quark mass ratios, such as 
$$|V_{cb}| \; \approx \; |V_{ts}| \; \approx \; \sqrt{2} ~
\left (\frac{m_s}{m_b}  - \frac{m_c}{m_t} \right ) \left [1 + 
3 \left (\frac{m_s}{m_b} + \frac{m_c}{m_t} \right ) \right ] \; ,$$
are obtainable from a simple {\it Ansatz} of flavor 
permutation symmetry breaking at the weak scale. 
We prescribe the same {\it Ansatz} at the supersymmetric grand unified 
theory scale, and find that its all low-energy consequences on flavor mixings 
and $CP$ violation are in good agreement with current experimental data.
\end{abstract}

\vspace{1cm}

\begin{center}
PACS number(s): 12.15.Ff, 11.30.Hv, 11.30.Pb, 12.10.Dm
\end{center}

\newpage

\section{Introduction}
\setcounter{equation}{0}

The discovery of the top quark at Fermilab fulfilled the three-family 
quark picture in the standard electroweak model.
Up to now, some knowledge on the mass spectra of $(u, c, t)$ and 
$(d, s, b)$ quarks has been accumulated through both experimental and
theoretical (or phenomenological) attempts \cite{Gasser}. 
The ratios of quark mass 
eigenvalues are obtainable after one renormalizes them
to a common reference scale, e.g., $\mu = 1$ GeV or $M_Z$. There exists 
a clear mass hierarchy in each quark sector:
\begin{equation}
m_u \; \ll \; m_c \; \ll \; m_t \; ; ~~~~~~~~
m_d \; \ll \; m_s \; \ll \; m_b \; .
\end{equation}
In comparison, the masses of three charged leptons manifest a similar
hierarchical pattern \cite{PDG}. 

\vspace{0.3cm}

Quark mass eigenstates are related to quark weak (flavor)
eigenstates by the Kobayashi-Maskawa (KM) matrix $V$ \cite{KM}, which
provides a quite natural description of flavor mixings and $CP$ violation
in the standard model. To date, many experimental constraints on
the magnitudes of the KM matrix elements have been achieved.
The unitarity of $V$ together with current data requires 
a unique hierarchy among the nine matrix elements \cite{XingV}:
\begin{eqnarray}
|V_{tb}| \; > \; |V_{ud}| \; > \; |V_{cs}| & \gg & |V_{us}| \; > \; |V_{cd}| \; 
\nonumber \\
& \gg & |V_{cb}| \; > \; |V_{ts}| \; \nonumber \\
& \gg & |V_{td}| \; > \; |V_{ub}| \; > \; 0 \; .
\end{eqnarray}
Here $|V_{ub}|\neq 0$ is a necessary condition for the presence of $CP$ 
violation in the KM matrix.
 
\vspace{0.3cm}

How to understand the hierarchies of quark masses and flavor mixings is an important
but unsolved problem in particle physics. A natural approach to the final solution
of this problem is to look for the most favorable pattern of quark mass matrices
(see, e.g., Refs. \cite{Weinberg,Fritzsch77}),
which can account for all low-energy phenomena of quark mixings and $CP$ violation. The 
relevant symmetries hidden in such phenomenological schemes are possible to provide
useful hints toward the dynamical details of fermion mass generation.

\vspace{0.3cm}

It has been speculated by some authors that the realistic fermion mass matrices 
could arise from the flavor permutation symmetry and its spontaneous or
explicit breaking \cite{Democracy,Fritzsch90,Meshkov}. 
Under exact $S(3)_{\rm L}\times S(3)_{\rm R}$ symmetry the
mass spectrum for either up or down quark sector consists of only two levels:
one is of 2-fold degeneracy with vanishing mass eigenvalues, and the other 
is nondegenerate (massive). An appropriate breakdown of the above symmetry may 
lead to the observed mass hierarchy and flavor mixings. 
Although the way to introduce the minimum number
of free parameters for permutation symmetry breaking is technically
trivial, its consequences on quark mixings and $CP$ violation 
may be physically instructive and may even shed some light on the proper 
relations between the KM matrix elements and quark mass ratios.
Indeed there has not been a satisfactory symmetry breaking pattern with enough
predictive power in the literature.

\vspace{0.3cm}

In this work we first stress that some observed properties of the KM
matrix can be interpreted by the quark mass hierarchy without the 
assumption of specific mass matrices. In the quark mass limits such as
$m_u=m_d=0$, $m_t\rightarrow \infty$ or $m_b\rightarrow \infty$, 
we find that simple but instructive relations between the KM matrix
elements and quark mass ratios are suggestible from current experimental data.
Then we present a new quark mass {\it Ansatz} through the explicit
breakdown of flavor permutation symmetry at the weak scale ($M_Z = 91.187$ GeV).
This {\it Ansatz} contains seven free parameters, thus it can give rise to three 
predictions for the phenomena of quark mixings and $CP$ violation. The typical
results are $|V_{cb}| \approx |V_{ts}| \approx \sqrt{2} ~ (m_s/m_b - m_c/m_t)$,
$|V_{ub}/V_{cb}|\approx \sqrt{m_u/m_c}$ and $|V_{td}/V_{ts}|\approx 
\sqrt{m_d/m_s}$ in the leading order approximation. 
Prescribing the same {\it Ansatz} at the supersymmetric grand unified theory 
(GUT) scale ($M_X = 10^{16}$ GeV),
we derive the renormalized quark mass matrices at $M_Z$ for
small $\tan\beta_{\rm susy}$ (the ratio of Higgs vacuum expectation values
in the minimal supersymmetric model). We also
renormalize some relations between the KM matrix elements and quark mass ratios
at $M_Z$ for arbitrary $\tan\beta_{\rm susy}$, and find that the relevant
results are in good agreement with experimental data.
The scale-independent predictions of our {\it Ansatz} for the characteristic
measurables of $CP$ asymmetries in weak $B$ decays, i.e., 
$0.18 \leq \sin(2\alpha) \leq 0.58$, $0.5\leq \sin(2\beta) \leq 0.78$
and $-0.08 \leq \sin(2\gamma) \leq 0.5$, can be tested at the forthcoming
KEK and SLAC $B$-meson factories.

\vspace{0.3cm}

The remaining part of this paper is organized as follows. 
Some qualitative implications of the quark mass hierarchy on the KM matrix
elements, which are almost independent of the specific forms of quark mass 
matrices, are discussed in section 2. In section 3 we suggest 
a new quark mass {\it Ansatz} from the flavor permutation symmetry breaking
at the weak scale, and study its various consequences on flavor mixings 
and $CP$ violation. The same {\it Ansatz} is prescribed at the supersymmetric
GUT scale in section 4. By use of the one-loop renormalization group
equations, we run the mass matrices from $M_X$ to $M_Z$
and then discuss the renormalized relations between the KM matrix elements and
quark mass ratios. Section 5 is devoted to a brief summary of this work.

\section{Flavor mixings in quark mass limits}
\setcounter{equation}{0}

Without loss of any generality, the up and down quark mass matrices (denoted
by $M_{\rm u}$ and $M_{\rm d}$, respectively) can be chosen to be Hermitian.
After the diagonalization of $M_{\rm u}$ and $M_{\rm d}$ through the unitary
transformations
\begin{eqnarray}
O^{\dagger}_{\rm u} M_{\rm u} O_{\rm u} & = & {\rm Diag}\{ m_u, ~ m_c, ~ m_t
\} \; , \nonumber \\
O^{\dagger}_{\rm d} M_{\rm d} O_{\rm d} & = & {\rm Diag}\{ m_d, ~ m_s, ~ m_b
\} \; ,
%
\end{eqnarray}
one obtains the KM matrix $V\equiv O^{\dagger}_{\rm u} O_{\rm d}$, which describes
quark flavor mixings in the charged current. Explicitly, the KM matrix elements
read
\begin{equation}
V_{ij} \; = \; \sum^3_{k=1} \left ( O^{{\rm u}^*}_{ki} ~ O^{\rm d}_{kj} \right ) \; ,
\end{equation}
depending upon the quark mass ratios $m_u/m_c$, $m_c/m_t$ (from $O_{\rm u}$) and
$m_d/m_s$, $m_s/m_b$ (from $O_{\rm d}$) as well as other parameters of $M_{\rm u}$
and $M_{\rm d}$ (e.g., the non-trivial phase shifts between $M_{\rm u}$ and $M_{\rm d}$).
In view of the distinctive mass hierarchy in Eq. (1.1), we find that some 
interesting properties of $V$ can be interpreted without the assumption of specific
forms of $M_{\rm u}$ and $M_{\rm d}$. 

\begin{center}
{\large\bf A. ~ $|V_{us}|$ and $|V_{cd}|$ in the limits $m_t\rightarrow \infty$ 
and $m_b\rightarrow \infty$}
\end{center}

Since the mass spectra of up and down quarks are absolutely dominated by
$m_t$ and $m_b$ respectively, the limits $m_t\rightarrow \infty$ and 
$m_b\rightarrow \infty$ are expected to be very reliable when we discuss flavor mixings
between $(u,d)$ and $(c,s)$. In this case, the effective mass matrices 
turn out to be two $2\times 2$ matrices and the resultant
flavor mixing matrix (i.e., the Cabibbo matrix \cite{Cabibbo}) 
cannot accommodate $CP$ violation. The magnitudes of $V_{us}$ and $V_{cd}$ can
be obtained from Eq. (2.2), since $O_{i3}=O_{3i}=\delta_{i3}$ holds for both
sectors in the above-mentioned mass limits. We find that
$|V_{us}| = |V_{cd}|$ 
is a straightforward result guaranteed by the unitarity of $O_{\rm u}$ and $O_{\rm d}$.
The current experimental data, together with unitary conditions of the
$3\times 3$ KM matrix, have implied \cite{XingV,XingW} 
\begin{equation}
|V_{us}| ~ - ~ |V_{cd}| \; \approx \; A^2\lambda^5 \left (\frac{1}{2} - \rho
\right ) \; < \; 10^{-3} \; ,
\end{equation}
which is insensitive to allowed errors of the Wolfenstein parameters
$A$, $\lambda$ and $\rho$ \cite{Wolfenstein}. From the discussions above we realize
that the near equality of $|V_{us}|$ and $|V_{cd}|$ is in fact
a natural consequence of $m_t\gg m_c, m_u$ and $m_b \gg m_s, m_d$.

\vspace{0.3cm}

The magnitude of $V_{us}$ (or $V_{cd}$) must be a function of the mass ratios 
$m_u/m_c$ and $m_d/m_s$ in the limits $m_t\rightarrow \infty$ and
$m_b\rightarrow \infty$, if $M_{\rm u}$ and $M_{\rm d}$ have 
parallel or quasi-parallel structures. Considering the experimentally allowed 
regions of $m_u/m_c$ ($\sim 5\times 10^{-3}$ \cite{PDG}), $m_s/m_d$ ($= 18.9\pm 0.8$
\cite{Leutwyler}) and $|V_{us}|$ ($=0.2205\pm 0.0018$ \cite{PDG}), 
one may guess that $|V_{us}|$ is dominated by
$\sqrt{m_d/m_s}$ but receives small correction from $\sqrt{m_u/m_c}$.
Indeed such an instructive result for $|V_{us}|$ or $|V_{cd}|$
can be derived from $2\times 2$ Hermitian mass matrices of the form \cite{Weinberg}
\begin{equation}
\left ( \matrix{
0	&	A \cr
A^*	&	B } \right ) \; ,
\end{equation}
where $|B|\gg |A|$. Denoting the phase difference between $A_{\rm u}$ and
$A_{\rm d}$ as $\Delta\phi$, we obtain
\begin{equation}
|V_{us}| \; = \; |V_{cd}| \; = \;
\left | ~ \sqrt{\frac{m_c}{m_u + m_c}} \sqrt{\frac{m_d}{m_d + m_s}}
~ - ~ \exp({\rm i} \Delta\phi) \sqrt{\frac{m_u}{m_u + m_c}} \sqrt{\frac{m_s}{m_d + m_s}}
~ \right | \; .
\end{equation}
Although the $2\times 2$ flavor mixing matrix cannot accommodate $CP$ violation,
the phase shift $\Delta\phi$ is non-trivial on the point that it sensitively determines
the value of $|V_{us}|$. For illustration, we calculate the allowed region of $\Delta\phi$
as a function of $m_u/m_c$ in Fig. 1. It is clear that the possibilities 
$\Delta\phi=0^0$, $90^0$ and $180^0$ have all been ruled out by current data 
on $V_{us}$ and $m_s/m_d$, since $m_u/m_c \geq 10^{-3}$ is expected to be true. 
We conclude that the presence of $\Delta\phi$ in the quark mass {\it Ansatz} above 
is crucial for correct reproduction of $|V_{us}|$ and $|V_{cd}|$.
Such a non-trivial phase shift will definitely lead to $CP$ violation,
when the limits $m_t\rightarrow \infty$ and $m_b\rightarrow \infty$ are discarded.

\begin{center}
{\large\bf B. ~ $|V_{cb}|$ and $|V_{ts}|$ in the limit $m_u=m_d=0$}
\end{center}

Considering the fact that $m_u$ and $m_d$ are negligibly small in
the mass spectra of up and down quarks, one can take the reasonable
limit $m_u=m_d=0$ to discuss flavor mixings between the second and 
third families. In this case, there is no mixing between $(u,d)$ and $(c,s)$
or between $(u,d)$ and $(t,b)$. Thus $M_{1i}=M_{i1}=0$ holds for both
up and down mass matrices, and then we get $O_{1i}=O_{i1}=\delta_{1i}$.
The relation $|V_{cb}| = |V_{ts}|$ is straightforwardly obtainable
from Eq. (2.2) by use of the unitary conditions of $O_{\rm u}$ and $O_{\rm d}$.
In contrast, the present data and unitarity of the KM matrix requires \cite{XingW}
\begin{equation}
|V_{cb}| ~ - ~ |V_{ts}| \; \approx \; A \lambda^4 \left (\frac{1}{2} - \rho
\right ) \; < \; 10^{-2} \; .
\end{equation}
We see that the near equality between $|V_{cb}|$ and $|V_{ts}|$ 
can be well understood, because the quark mass limit $m_u=m_d=0$ is 
a good approximation for $M_{\rm u}$ and $M_{\rm d}$.

\vspace{0.3cm}

We expect that $|V_{cb}|$ and $|V_{ts}|$ are functions of the mass ratios
$m_c/m_t$ and $m_s/m_b$ in the limit $m_u=m_d=0$. Current experimental
data give $|V_{cb}|=0.0388\pm 0.0032$ \cite{Neubert}, while 
$m_c/m_t \sim 10^{-3}$ \cite{PDG} and $m_b/m_s = 34\pm 4$ \cite{Narison}
are allowed. Thus $|V_{cb}|$ (or $|V_{ts}|$) should be dominated by
$m_s/m_b$, and it may get a little correction from $m_c/m_t$. To obtain
a linear relation among $V_{cb}$, $m_s/m_b$ and $m_c/m_t$ in the leading order
approximation, one can investigate mass matrices of the following Hermitian form:
\begin{equation}
\left (\matrix{
0	& 0	& 0 \cr
0	& A	& B \cr
0	& B^*	& C } \right ) \; ,
\end{equation}
where $A\neq 0$ and $|C|\gg |B| \sim |A|$ for both quark sectors.
This generic pattern can also be regarded as a trivial generalization of the
Fritzsch {\it Ansatz}, in which $A=0$ is assumed \cite{Fritzsch77},
but they have rather different consequences on
the magnitudes of $V_{cb}$ and $V_{ts}$.
Denoting $\Delta\varphi = \arg(B_{\rm u}/B_{\rm d})$, 
$R_{\rm u} = |B_{\rm u}/A_{\rm u}|$ and $R_{\rm d} = |B_{\rm d}/
A_{\rm d}|$, we find the approximate result
\begin{equation}
|V_{cb}| \; = \; |V_{ts}| \; \approx \; 
\left | ~ R_{\rm d} \frac{m_s}{m_b} ~ - ~ \exp({\rm i} \Delta\varphi)
R_{\rm u} \frac{m_c}{m_t} ~ \right | \; .
\end{equation}
One can see that $|V_{cb}|\propto m_s/m_b$ holds approximately, if
$R_{\rm u}$  is comparable in magnitude with $R_{\rm d}$.
Here the phase shift $\Delta\varphi$ plays an insignificant
(negligible) role in confronting Eq. (2.8) with the experimental
data on $|V_{cb}|$, since the term proportional to $m_c/m_t$ is
significantly suppressed. To determine the values of $R_{\rm u}$
and $R_{\rm d}$, however, one has to rely on a more specific
{\it Ansatz} of quark mass matrices. 

\begin{center}
{\large\bf C. ~ $|V_{ub}/V_{cb}|$ in $m_b\rightarrow \infty$
and $|V_{td}/V_{ts}|$ in $m_t\rightarrow \infty$}
\end{center}

Now let us take a look at the two smallest matrix elements of $V$,
$|V_{ub}|$ and $|V_{td}|$, in the quark mass limits.
Taking $m_b\rightarrow \infty$, we have $O^{\rm d}_{i3}=
O^{\rm d}_{3i}=\delta_{i3}$, because $M_{\rm d}$ turns out to be
an effective $2\times 2$ matrix in this limit. Then the ratio of
$|V_{ub}|$ to $|V_{cb}|$ reads
\begin{equation}
\lim_{m_b\rightarrow \infty} \left | \frac{V_{ub}}{V_{cb}} \right |
\; = \; \left | \frac{O^{\rm u}_{31}}{O^{\rm u}_{32}} \right | \; ,
\end{equation}
obtained from Eq. (2.2). Contrary to common belief, $|V_{ub}/V_{cb}|$ is absolutely
independent of the mass ratio $m_d/m_s$ in the limit $m_b\rightarrow
\infty$! Therefore one expects that the left-handed side of Eq. (2.9) is 
dominated by a simple function of the mass ratio $m_u/m_c$, while the
contribution from $m_c/m_t$ should be insignificant in most cases.
The present numerical knowledge of $|V_{ub}/V_{cb}|$ ($=0.08\pm 0.02$
\cite{PDG}) and $m_u/m_c$ ($\sim 5\times 10^{-3}$ \cite{PDG}) implies
that $|V_{ub}/V_{cb}|\approx \sqrt{m_u/m_c}$ is likely to be true.
Indeed such an approximate result can be reproduced from
the Fritzsch {\it Ansatz} and a variety of its modified versions
\cite{Hall}.

\vspace{0.3cm}

In the mass limit $m_t\rightarrow \infty$, $M_{\rm u}$ becomes an effective
$2\times 2$ matrix, and then $O^{\rm u}_{i3}=O^{\rm u}_{3i}=\delta_{i3}$
holds. The ratio of $|V_{td}|$ to $|V_{ts}|$ is obtainable from Eq. (2.2)
as follows:
\begin{equation}
\lim_{m_t\rightarrow \infty} \left | \frac{V_{td}}{V_{ts}} \right |
\; = \; \left | \frac{O^{\rm d}_{31}}{O^{\rm d}_{32}} \right | \; .
\end{equation}
Here again we find that $|V_{td}/V_{ts}|$ is independent of both
$m_u/m_c$ and $m_c/m_t$ in the limit $m_t\rightarrow \infty$, thus it
may be a simple function of the mass ratios $m_d/m_s$ and $m_s/m_b$.
The current data give $0.15 \leq |V_{td}/V_{ts}| \leq 0.34$ \cite{Ali},
$m_s/m_d = 18.9\pm 0.8$ \cite{Leutwyler} and $m_b/m_s =34\pm 4$ \cite{Narison}.
We expect that $|V_{td}/V_{ts}|\approx \sqrt{m_d/m_s}$ has a 
large chance to be true in the leading order approximation. 
Note that this approximate relation can also be derived from the 
Fritzsch {\it Ansatz} or some of its revised versions \cite{Hall}.

\vspace{0.3cm}

The qualitative discussions above have shown that
some properties of the KM matrix $V$ can be well understood just
from the quark mass hierarchy. For example, $|V_{us}|\approx |V_{cd}|$
and $|V_{cb}|\approx |V_{ts}|$ are natural consequences of arbitrary
(Hermitian) mass matrices with $m_3 \gg m_2, m_1$ and $m_1 \ll m_2, m_3$ 
respectively, where $m_i$ denote the mass eigenvalues of each quark
sector. To a good degree of accuracy, $|V_{us}|$ and $|V_{cd}|$ are
expected to be 
independent of the mass ratios $m_c/m_t$ and $m_s/m_b$, while
$|V_{cb}|$ and $|V_{ts}|$ are independent of $m_u/m_c$ and $m_d/m_s$.
The ratios $|V_{ub}/V_{cb}|$ and $|V_{td}/V_{ts}|$ may be simple functions 
of $m_u/m_c$ and $m_d/m_s$, respectively, in the leading order approximations.
These qualitative results should hold, in most cases and without fine
tuning effects, for generic (Hermitian) forms of $M_{\rm u}$ and $M_{\rm d}$.
They can be used as an enlightening clue for the construction of specific
and predictive $Ans\ddot{a}tze$ of quark mass matrices.

\section{A quark mass {\it Ansatz} at the weak scale}
\setcounter{equation}{0}

We are now in a position to consider the realistic $3\times 3$ 
mass matrices in no assumption of the quark mass limits. Such
an {\it Ansatz} should be able to yield the definite 
values of $R_{\rm u}$ and $R_{\rm d}$ in Eq. (2.8), and account for
current experimental data on flavor mixings and $CP$ violation 
at low-energy scales. 

\begin{center}
{\large\bf A. ~ Flavor permutation symmetry breaking}
\end{center}

We start from the flavor permutation symmetry to construct 
quark mass matrices at the weak scale, so that the resultant
KM matrix can be directly confronted with the experimental data.
The mass matrix with the $S(3)_{\rm L}\times S(3)_{\rm R}$ symmetry
reads
\begin{equation}
M_0 \; = \; \frac{c}{3} \left ( \matrix{
1	& 1	& 1 \cr
1	& 1	& 1 \cr
1	& 1	& 1 } \right ) \; ,
\end{equation}
where $c=m_3$ denotes the mass eigenvalue of the third-family
quark ($t$ or $b$). Note that $M_0$ is obtainable from another
rank-one matrix 
\begin{equation}
M_{\rm H} \; = \; c \left ( \matrix{
0	& 0	& 0 \cr
0	& 0	& 0 \cr
0	& 0	& 1 } \right ) \;
\end{equation}
through the unitary transformation $M_0 = U^{\dagger} M_{\rm H} U$,
where
\begin{equation}
U \; = \; \frac{1}{\sqrt{6}} \left ( \matrix{
\sqrt{3}	& -\sqrt{3}	& 0 \cr
1	& 1	& -2 \cr
\sqrt{2}	& \sqrt{2}	& \sqrt{2} } \right ) \; .
\end{equation}
To generate masses for the second- and first-family quarks, one has to
break the permutation symmetry of $M_0$ to
the $S(2)_{\rm L}\times S(2)_{\rm R}$ and $S(1)_{\rm L}\times S(1)_{\rm R}$
symmetries, respectively. 
Here we assume that the up and down mass 
matrices have the parallel symmetry breaking patterns, corresponding to
the parallel dynamical details of quark mass generation. We further assume
that each symmetry breaking chain (i.e., $S(3)_{\rm L}\times S(3)_{\rm R}
\rightarrow S(2)_{\rm L}\times S(2)_{\rm R}$ or 
$S(2)_{\rm L}\times S(2)_{\rm R} \rightarrow S(1)_{\rm L}\times S(1)_{\rm R}$) 
is induced by a single real parameter,
and the possible phase shift between two quark sectors arises from
an unknown dynamical mechanism. 

\vspace{0.3cm}

With the assumptions made above, a new {\it Ansatz} for the up and down
mass matrices can be given as follows:
\begin{equation}
M^{\prime}_0 \; = \; \frac{c}{3} \left [ 
\left ( \matrix{
1	& 1	& 1 \cr
1	& 1	& 1 \cr
1	& 1	& 1 } \right ) 
+ \epsilon \left ( \matrix{
0	& 0	& 1 \cr
0	& 0	& 1 \cr
1	& 1	& 1 } \right ) 
+ \sigma \left ( \matrix{
1	& 0	& -1 \cr
0	& -1	& 1 \cr
-1	& 1	& 0 } \right ) \right ] \; ,
\end{equation}
where $\epsilon$ and $\sigma$ are real (dimensionless) perturbation
parameters responsible for the breakdowns of $S(3)_{\rm L}\times
S(3)_{\rm R}$ and $S(2)_{\rm L}\times S(2)_{\rm R}$ symmetries of
$M_0$, respectively. In the basis of $M_{\rm H}$, the mass matrix 
$M^{\prime}_0$ takes the form
\begin{equation}
M^{\prime}_{\rm H} \; = \; c
\left (\matrix{
0	& \displaystyle\frac{\sqrt{3}}{3} \sigma	& 0 \cr \cr
\displaystyle\frac{\sqrt{3}}{3} \sigma 
& \displaystyle -\frac{2}{9} \epsilon
& \displaystyle -\frac{2\sqrt{2}}{9} \epsilon \cr \cr
0	& \displaystyle -\frac{2\sqrt{2}}{9} \epsilon
& \displaystyle 1+\frac{5}{9} \epsilon \cr } \right ) \; ,
\end{equation}
which has three free parameters and three texture zeros.
Diagonalizing $M^{\prime}_{\rm H}$ through the unitary
transformation $O^{{\prime}^{\dagger}} M^{\prime}_{\rm H} O^{\prime}
= {\rm Diag} \{ m_1, ~ m_2, ~ m_3 \}$, 
one can determine $c$, $\epsilon$
and $\sigma$ in terms of the quark mass eigenvalues. In the
next-to-leading order approximations, we get
\begin{eqnarray}
c & \approx & m_3 \left ( 1 + \frac{5}{2} \frac{m_2}{m_3}
\right ) \; , \nonumber \\
\epsilon & \approx & -\frac{9}{2} \frac{m_2}{m_3}
\left ( 1- \frac{1}{2} \frac{m_2}{m_3} \right ) \; , \nonumber \\
\sigma & \approx & \frac{\sqrt{3 m_1 m_2}}{m_3} 
\left ( 1 - \frac{5}{2} \frac{m_2}{m_3} \right ) \; .
\end{eqnarray}
Then the elements of $O^{\prime}$ are expressible in terms of
the mass ratios $m_1/m_2$ and $m_2/m_3$.

\vspace{0.3cm}

The flavor mixing matrix can be given as $V= O^{{\prime}^{\dagger}}_{\rm u}
P O^{\prime}_{\rm d}$, where $P$ is a diagonal phase matrix taking
the form $P={\rm Diag} \{ 1, ~ \exp({\rm i} \Delta \phi), 
~ \exp({\rm i} \Delta \phi) \}$.
Here $\Delta \phi$ denotes the phase shift between up and
down mass matrices, and its presence is necessary for the {\it Ansatz}
to correctly reproduce both $|V_{us}|$ (or $|V_{cd}|$) and $CP$ violation.

\begin{center}
{\large\bf B. ~ Flavor mixings and $CP$ violation}
\end{center}

Calculating the KM matrix elements $|V_{us}|$ and $|V_{cd}|$ 
in the next-to-leading order approximation, we obtain 
\begin{equation}
|V_{us}| \; \approx \; |V_{cd}| \; \approx \; 
\sqrt{\left (\frac{m_u}{m_c} +\frac{m_d}{m_s} - 2 \sqrt{\frac{m_u m_d}{m_c m_s}}
~ \cos \Delta\phi \right )
\left ( 1-\frac{m_u}{m_c}-\frac{m_d}{m_s}\right )} \; .
\end{equation}
This result is clearly consistent with that in Eq. (2.5). The allowed 
region of $\Delta\phi$ has been shown by Fig. 1 with the inputs of 
$m_s/m_d$ and $|V_{us}|$. We find $73^0 \leq \Delta\phi \leq 82^0$
for reasonable values of $m_u/m_c$. In the leading order
approximation of Eq. (3.7) or Eq. (2.5), it is easy to check that
$|V_{cd}|$, $\sqrt{m_u/m_c}$ and $\sqrt{m_d/m_s}$ form a triangle in
the complex plane \cite{FritzschXing}. 
 
\vspace{0.3cm}

In the next-to-leading order approximation, $|V_{cb}|$ and $|V_{ts}|$
can be given as
\begin{equation}
|V_{cb}| \; \approx \; |V_{ts}| \; \approx \; \sqrt{2}
\left (\frac{m_s}{m_b} -\frac{m_c}{m_t}\right ) 
\left [ 1+3 \left (\frac{m_s}{m_b} + \frac{m_c}{m_t}\right ) \right ] \; .
\end{equation}
Comparing between Eqs. (3.8) and (2.8), we get 
$R_{\rm u} = R_{\rm d} = \sqrt{2}$, determined by the
quark mass {\it Ansatz} in Eq. (3.4).
By use of $m_b/m_s = 34\pm 4$ \cite{Narison}, we illustrate the allowed
region of $|V_{cb}|$ as a function of $m_c/m_t$ in Fig. 2, where
the experimental constraint on $|V_{cb}|$ ($=0.0388 \pm 0.0032$ 
\cite{Neubert})
has also been shown. We see that the result of $|V_{cb}|$
obtained in Eq. (3.8) is rather favored by current data. This implies
that the pattern of permutation symmetry breaking 
(i.e., $S(3)_{\rm L}\times S(3)_{\rm R} \rightarrow S(2)_{\rm L}
\times S(2)_{\rm R}$) in Eq. (3.4) may have a large chance to be true.

\vspace{0.3cm}

The ratios $|V_{ub}/V_{cb}|$ and $|V_{td}/V_{ts}|$ are found to be
\begin{equation}
\left | \frac{V_{ub}}{V_{cb}} \right | \; \approx \; \sqrt{\frac{m_u}{m_c}} \; ,
~~~~~~~~
\left | \frac{V_{td}}{V_{ts}} \right | \; \approx \; \sqrt{\frac{m_d}{m_s}} \; 
\end{equation}
to a good degree of accuracy 
\footnote{More exactly, we obtain 
$|V_{ub}/V_{cb}| \approx \sqrt{m_u/m_c} ~ (1 - \delta )$ with
$\delta = \sqrt{(m_c m_d)/(m_u m_s)} ~ (m_s/m_b) \cos\Delta\phi$.
The magnitude of $\delta$ may be as large as $10\%$ to $15\%$
for $\Delta\phi \sim 0^0$ or $180^0$, but it is only about $2\%$ 
for $73^0 \leq \Delta\phi \leq 82^0$ allowed by Eq. (3.7).}.
By use of Leutwyler's result $m_s/m_d = 18.9\pm 0.8$ \cite{Leutwyler}, we get 
$0.225\leq |V_{td}/V_{ts}| \leq 0.235$.
In comparison, the current data together with unitarity of 
the $3\times 3$ KM matrix yield $0.15\leq |V_{td}/V_{ts}|\leq 0.34$ 
\cite{Ali}. The allowed region of 
$|V_{ub}/V_{cb}|$ is constrained by that of $m_u/m_c$, which has not been
reliably determined. 
We find that $0.0036 \leq m_u/m_c \leq 0.01$ is necessary for the quark mass 
{\it Ansatz} in Eq. (3.4) to accommodate the experimental
result $|V_{ub}/V_{cb}| = 0.08\pm 0.02$ \cite{PDG}. 

\vspace{0.3cm}

In the leading order approximations, we have
$|V_{ud}|\approx |V_{cs}| \approx |V_{tb}| \approx 1$. Small corrections to
these diagonal elements are obtainable with the help of the unitary
conditions of $V$. If we rescale three sides of the unitarity triangle
$V^*_{ub}V_{ud} + V^*_{cb}V_{cd} + V^*_{tb}V_{td} =0$ by $V^*_{cb}$,
then the resultant triangle is approximately equivalent to that formed
by $V_{cd}$, $\sqrt{m_u/m_c}$ and $\sqrt{m_d/m_s}$ in the complex plane
\cite{FritzschXing}. This interesting result can be easily shown by use of 
Eqs. (3.7), (3.8) and (3.9). Three inner angles of the unitarity triangle 
turn out to be
\begin{eqnarray}
\alpha & = & \arg \left (- \frac{V^*_{ub}V_{ud}}{V^*_{tb}V_{td}} \right )
\; \approx \; \Delta\phi \; , \nonumber \\
\beta & = & \arg \left (- \frac{V^*_{tb}V_{td}}{V^*_{cb}V_{cd}} \right )
\; \approx \; \tan \left (\frac{\sin\Delta \phi}
{\displaystyle \sqrt{\frac{m_c m_d}{m_u m_s}} - \cos\Delta \phi} \right ) \; ,
\nonumber \\
\gamma & = & \arg \left (- \frac{V^*_{cb}V_{cd}}{V^*_{ub}V_{ud}} \right )
\; \approx \; 180^0 - \alpha - \beta \; 
\end{eqnarray}
in the approximations made above. At the forthcoming $B$-meson factories,
these three angles will be determined from $CP$ asymmetries in 
a variety of weak $B$ decays (e.g., $B_d \rightarrow J/\psi K_S$,
$B_d\rightarrow \pi^+\pi^-$ and $B_s\rightarrow \rho^0 K_S$). 
For illustration, we calculate
$\sin (2\alpha)$, $\sin (2\beta)$ and $\sin (2\gamma)$ by use of Eq. (3.10)
and plot their allowed regions in Fig. 3. 
Clearly the quark mass {\it Ansatz} under discussion favors 
$0.18\leq \sin(2\alpha) \leq 0.58$, $0.5\leq \sin(2\beta) \leq 0.78$
and $-0.08 \leq \sin(2\gamma) \leq 0.5$. These results do not involve
large errors, and they can be confronted with the relevant experiments
of $B$ decays and $CP$ violation in the near future.

\vspace{0.3cm}

Finally we point out that $CP$ violation in the KM matrix, measured by the
Jarlskog parameter $J$ \cite{Jarlskog}, can also be estimated in terms of quark
mass ratios. It is easy to obtain
\begin{equation}
J \; \approx \; 2 \sqrt{\frac{m_u}{m_c}} \sqrt{\frac{m_d}{m_s}}
\left ( \frac{m_s}{m_b} - \frac{m_c}{m_t} \right )^2 
\left [ 1 + 6 \left ( \frac{m_s}{m_b} + \frac{m_c}{m_t} \right ) \right ] 
\sin\Delta\phi \; .
\end{equation}
Typically taking $m_u/m_c = 0.005$, $m_s/m_d = 19$, $m_c/m_t = 0.005$,
$m_b/m_s = 34$ and $\Delta\phi = 80^0$, we get $J\approx 2.3 \times 10^{-5}$. 
This result is of course consistent with current data on $CP$ violation in the 
$K^0-\bar{K}^0$ mixing system \cite{PDG}.

\section{A quark mass {\it Ansatz} at the GUT scale}
\setcounter{equation}{0}

It is interesting to speculate that the quark mass hierarchy and flavor
mixings may arise from a certain symmetry breaking pattern in the
context of supersymmetric GUTs \cite{SUSY,Froggatt}. 
Starting from the flavor permutation
symmetry, here we prescribe the same {\it Ansatz} for quark mass matrices
as that proposed in Eq. (3.4) at the supersymmetric GUT scale $M_X$.
For simplicity we use $\hat{M}_0$ and $\hat{M}_{\rm H}$,
which correspond to $M_0^{\prime}$ in Eq. (3.4) and $M^{\prime}_{\rm H}$
in Eq. (3.5), to denote the mass matrices at $M_X$ in two
different bases. They are related to each other through the unitary
transformation $\hat{M}_0 = U^{\dagger} \hat{M}_{\rm H}
U$. The flavor mixing matrix derived from $\hat{M}_0$ (or
$\hat{M}_{\rm H}$) is denoted by $\hat{V}$. The subsequent 
running effects of $\hat{M}_0$ and $\hat{V}$ from $M_X$ to $M_Z$ can be
calculated with the help of the renormalization group equations in the
minimal supersymmetric standard model. 

\begin{center}
{\large\bf A. ~ Renormalized mass matrices at $M_Z$}
\end{center}

The simplicity of $\hat{M}_0$ (or $\hat{M}_{\rm H}$)
may be spoiled after it evolves from $M_X$ to $M_Z$. To illustrate 
this point, here we derive the renormalized mass matrices 
$\hat{M}^{\rm u}_0$ and $\hat{M}^{\rm d}_0$ at $M_Z$
by use of the one-loop renormalization group equations for the
Yukawa matrices and gauge couplings \cite{Babu}. 
To get instructive analytical
results, we constrain the ratio of Higgs vacuum expectation values 
$\tan\beta_{\rm susy}$ to be small enough ($\tan\beta_{\rm susy} < 10$), 
so that all non-leading
terms in the Yukawa couplings different from that of the top quark 
can be safely neglected \cite{Giudice}. 
In this approximation, the evolution equations
of $\hat{M}^{\rm u}_0$ and $\hat{M}^{\rm d}_0$ read
\begin{eqnarray}
16 \pi^2 \frac{{\rm d} \hat{M}^{\rm u}_0}{{\rm d} \chi}
& = & \left [ \frac{3}{v^2} {\rm Tr} \left ( \hat{M}^{\rm u}_0
\hat{M}^{{\rm u}^{\dagger}}_0 \right ) + \frac{3}{v^2} \left ( 
\hat{M}^{\rm u}_0 \hat{M}^{{\rm u}^{\dagger}}_0
\right ) - G_{\rm u} \right ] \hat{M}^{\rm u}_0 \; , \nonumber \\
16 \pi^2 \frac{{\rm d} \hat{M}^{\rm d}_0}{{\rm d} \chi}
& = & \left [ \frac{1}{v^2} \left ( 
\hat{M}^{\rm u}_0 \hat{M}^{{\rm u}^{\dagger}}_0
\right ) - G_{\rm d} \right ] \hat{M}^{\rm d}_0 \; , 
\end{eqnarray}
where $\chi = \ln (\mu /M_Z)$, 
$G_{\rm u}$ and $G_{\rm d}$ are functions of the gauge couplings
$g^{~}_i$ ($i=1,2,3$), and $v$ is the overall Higgs vacuum expectation
value normalized to 175 GeV. For the charged lepton mass matrix 
$\hat{M}^{\rm e}_0$, its evolution equation is dominated only by 
a linear term $G_{\rm e}$ in the case of small $\tan\beta_{\rm susy}$.
Thus the Hermitian structure of $\hat{M}^{\rm e}_0$ will be unchanged 
through the running from $M_X$ to $M_Z$ (in our discussions the neutrinos
are assumed to be massless). The quantity $G_{\rm n}$ 
(n = u, d or e) obeys the following equation:
\begin{equation}
G_{\rm n} (\chi) \; = \; 8\pi^2 \sum^3_{i=1} \left [ \frac{c^{\rm n}_i ~
g^2_i (0)}{8\pi^2 - b_i ~ g^2_i (0) ~ \chi} \right ] \; ,
\end{equation}
where $c^{\rm n}_i$ and $b_i$ are coefficients in the context of 
the minimal supersymmetric standard model. The values of $g^2_i (0)$,
$c^{\rm n}_i$ and $b_i$ are listed in Table 1.
\begin{table}
\begin{center}
\begin{tabular}{c|ccccc} \hline\hline 
$i$	& ~~ $c^{\rm u}_i$ ~~ 	& ~ $c^{\rm d}_i$ ~
	& ~~ $c^{\rm e}_i$ ~~	& ~~ $b_i$ ~~	& ~~ $g^2_i(0)$  \\ \hline  \\
1	& 13/9		& 7/9		& 3 	& 11	& 0.127 \\ \\
2	& 3	& 3	& 3	& 1	& 0.42 \\ \\
3	& 16/3		& 16/3		& 0	& $-$3	& 1.44 
\\ \\ \hline\hline
\end{tabular}
\end{center}
\caption{The values of $c^{\rm n}_i$, $b_i$ and $g^2_i(0)$ in the 
minimal supersymmetric standard model.}
\end{table}
In order to solve Eq. (4.1), we diagonalize $\hat{M}^{\rm u}_0$ through
the unitary transformation $\hat{O}^{\dagger} \hat{M}^{\rm u}_0 \hat{O}
= \hat{M}^{{\rm u}^{\prime}}_0$. Making the same transformation for
$\hat{M}^{\rm d}_0$, i.e., $\hat{O}^{\dagger} \hat{M}^{\rm d}_0 \hat{O}
= \hat{M}^{{\rm d}^{\prime}}_0$, we obtain the simplified evolution equations
as follows:
\begin{eqnarray}
16 \pi^2 \frac{{\rm d}\hat{M}^{{\rm u}^{\prime}}_0}{{\rm d} \chi}
& = & \left [ 3 f^2_t \left ( \matrix{
0	& 0	& 0 \cr
0	& 0	& 0 \cr
0	& 0	& 1 } \right ) +
\left ( 3 f^2_t - G_{\rm u} \right ) \left ( \matrix{
1	& 0	& 0 \cr
0	& 1	& 0 \cr
0	& 0	& 1 } \right ) \right ] \hat{M}^{{\rm u}^{\prime}}_0
\; , \nonumber \\
16 \pi^2 \frac{{\rm d}\hat{M}^{{\rm d}^{\prime}}_0}{{\rm d} \chi}
& = & \left [ f^2_t \left ( \matrix{
0	& 0	& 0 \cr
0	& 0	& 0 \cr
0	& 0	& 1 } \right ) -
G_{\rm u} \left ( \matrix{
1	& 0	& 0 \cr
0	& 1	& 0 \cr
0	& 0	& 1 } \right ) \right ] \hat{M}^{{\rm d}^{\prime}}_0 \; ,
\end{eqnarray}
where $f_t = m_t/v$ is the top quark Yukawa coupling eigenvalue. 
For simplicity in presenting the results, we define 
\begin{eqnarray}
\Omega_{\rm n} & = & \exp \left [ + \frac{1}{16\pi^2} \int^{\ln (M_X/M_Z)}_0
G_{\rm n}(\chi) ~ {\rm d}\chi \right ] \; , \nonumber \\
\xi_i & = & \exp \left [ - \frac{1}{16\pi^2} \int^{\ln (M_X/M_Z)}_0
f^2_i (\chi) ~ {\rm d}\chi \right ] 
\end{eqnarray}
with $i=t$ (or $i=b$). By use of Eq. (4.2) and the inputs listed in Table 1, one can 
explicitly calculate the magnitude of $\Omega_{\rm n}$. We find
$\Omega_{\rm u}=3.47$, $\Omega_{\rm d}=3.38$ and $\Omega_{\rm e}=1.49$
for $M_X=10^{16}$ GeV and $M_Z=91.187$ GeV.
The size of $\xi_t$ depends upon the value of $\tan\beta_{\rm susy}$
and will be estimated in the next subsection.
Solving Eq. (4.3) and transforming $\hat{M}^{{\rm n}^{\prime}}_0$
back to $\hat{M}^{\rm n}_0$, we get
\begin{eqnarray}
\hat{M}^{\rm u}_0 (M_Z) & = & \Omega_{\rm u} ~ \xi^3_t ~ \hat{O}
\left ( \matrix{
1	& 0	& 0 \cr
0	& 1	& 0 \cr
0	& 0	& \xi^3_t } \right )
\hat{O}^{\dagger} ~ \hat{M}^{\rm u}_0 (M_X) \; , \nonumber \\
\hat{M}^{\rm d}_0 (M_Z) & = & \Omega_{\rm d} ~ \hat{O}
\left ( \matrix{
1	& 0	& 0 \cr
0	& 1	& 0 \cr
0	& 0	& \xi_t  } \right )
\hat{O}^{\dagger} ~ \hat{M}^{\rm d}_0 (M_X) \; 
\end{eqnarray}
in the leading order approximation. 

\vspace{0.3cm}

Since $\hat{O}$ can be easily determined from $\hat{M}^{\rm u}_0 (M_X)$ and
$\hat{M}^{{\rm u}^{\prime}}_0 (M_X)$ in the approximation of
$\hat{M}^{{\rm u}^{\prime}}_0 (M_X) \approx {\rm Diag} \{0, ~ 0, ~ m_t \}$
made above,
we explicitly express $\hat{M}^{\rm u}_0 (M_Z)$ and $\hat{M}^{\rm d}_0 (M_Z)$
as follows:
\begin{equation}
\hat{M}^{\rm u}_0 (M_Z) \; = \; \frac{c_{\rm u}}{3} \Omega_{\rm u} \xi_t^3
\left [ \xi^3_t 
\left ( \matrix{
1	& 1	& 1 \cr
1	& 1	& 1 \cr
1	& 1	& 1 } \right ) 
+ \epsilon_{\rm u} \left ( \matrix{
x_{\rm u}	& x_{\rm u}	& y_{\rm u} \cr
x_{\rm u}	& x_{\rm u}	& y_{\rm u} \cr
y_{\rm u}	& y_{\rm u}	& z_{\rm u} } \right ) 
+ \sigma_{\rm u} \left ( \matrix{
1	& 0	& -1 \cr
0	& -1	& 1 \cr
-1	& 1	& 0 } \right ) \right ] \; 
\end{equation}
with $x_{\rm u} = (\xi^3_t -1)/9$, $y_{\rm u} =(7 \xi^3_t +2)/9$
and $z_{\rm u} = (13 \xi^3_t -4)/9$; and
\begin{equation}
\hat{M}^{\rm d}_0 (M_Z) \; = \; \frac{c_{\rm d}}{3} \Omega_{\rm d} 
\left [ \xi_t 
\left ( \matrix{
1	& 1	& 1 \cr
1	& 1	& 1 \cr
1	& 1	& 1 } \right ) 
+ \epsilon_{\rm d} \left ( \matrix{
x_{\rm d}	& x_{\rm d}	& y_{\rm d} \cr
x_{\rm d}	& x_{\rm d}	& y_{\rm d} \cr
y_{\rm d}	& y_{\rm d}	& z_{\rm d} } \right ) 
+ D_{\epsilon} 
+ \sigma_{\rm d} \left ( \matrix{
1	& 0	& -1 \cr
0	& -1	& 1 \cr
-1	& 1	& 0 } \right ) \right ] \; 
\end{equation}
with $x_{\rm d} = (\xi_t -1)/9$, $y_{\rm d} =(7 \xi_t +2)/9$,
$z_{\rm d} = (13 \xi_t -4)/9$ and
\begin{equation}
D_{\epsilon} \; = \; 2 \left (\epsilon_{\rm d} - \epsilon_{\rm u} \right )
x_{\rm d} \left ( \matrix{
1	& 1	& 1 \cr
1	& 1 	& 1 \cr
-2	& -2	& -2 } \right ) \; .
\end{equation}
If one takes $M_Z=M_X$, which leads to $\Omega_{\rm n}=1$,
$\xi_i =1$ and in turn $x_{\rm n}=0$, $y_{\rm n}=1$,
$z_{\rm n}=1$ and $D_{\epsilon}=0$,
then Eqs. (4.6) and (4.7) recover the form of 
$\hat{M}_0 (M_X)$ as assumed in Eq. (3.4).
To a good degree of accuracy, $\hat{M}^{\rm u}_0 (M_Z)$ 
remains Hermitian. The Hermiticity of $\hat{M}^{\rm d}_0 (M_Z)$
is violated by $D_{\epsilon}$, which would vanish
if the top and bottom quark masses were identical
(i.e., $\epsilon_{\rm d} = \epsilon_{\rm u}$). The presence of
nonvanishing $D_{\epsilon}$ reflects the fact that $m_t$ dominates
the mass spectra of both quark sectors \cite{Albright}.
Of course, one can transform the mass matrices obtained in Eqs. (4.6)
and (4.7) into 
the basis of $\hat{M}_{\rm H}$. In doing so, we will find the inequality
between (2,3) and (3,2) elements of $\hat{M}^{\rm d}_{\rm H} (M_Z)$,
arising from $D_{\epsilon}$.

\begin{center}
{\large\bf B. ~ Renormalized flavor mixings at $M_Z$}
\end{center}

Calculating the magnitudes of flavor mixings from $\hat{M}_0$ or 
$\hat{M}_{\rm H}$ at
$M_X$, we can obtain the same asymptotic relations between the KM matrix 
elements and quark mass ratios as those given in Eqs. (3.7), (3.8) and (3.9). 
Now we renormalize such relations at the weak scale $M_Z$ by 
means of the renormalization group equations. The quantities 
$\xi_t$ and $\xi_b$ defined in Eq. (4.4) will be evaluated below for arbitrary 
$\tan\beta_{\rm susy}$, 
so that one can get some quantitative feeling about the running effects
of quark mass matrices and flavor mixings from $M_X$ to $M_Z$.

\vspace{0.3cm}

The one-loop renormalization group equations for quark mass ratios and
elements of the KM matrix $\hat{V}$ have been explicitly presented by 
Babu and Shafi in Ref. \cite{Babu}.
In view of the hierarchy of Yukawa couplings and quark mixing angles, one can 
make reliable analytical approximations for the relevant evolution equations 
by keeping only the leading terms. It has been found that 
(1) the running effects of $m_u/m_c$ and $m_d/m_s$ are negligibly small;
(2) the diagonal elements of the KM matrix have negligible evolutions with energy;
(3) the evolutions of $|\hat{V}_{us}|$ and $|\hat{V}_{cd}|$ involve 
the second-family Yukawa couplings and thus they are negligible; 
(4) the KM matrix elements $|\hat{V}_{ub}|$, $|\hat{V}_{cb}|$, $|\hat{V}_{td}|$ 
and $|\hat{V}_{ts}|$ have identical running behaviors. 
Considering these points as well as the dominance of the third-family 
Yukawa couplings (i.e., $f_t$ and $f_b$), we get three key evolution equations 
in the minimal supersymmetric standard model:
\begin{eqnarray}
\left . \frac{m_s}{m_b} \right |_{M_Z} & = & \frac{1}{\xi_t ~ \xi^3_b} ~
\left . \frac{m_s}{m_b} \right |_{M_X} \; , \nonumber \\
\left . \frac{m_c}{m_t} \right |_{M_Z} & = & \frac{1}{\xi^3_t ~ \xi_b} ~ 
\left . \frac{m_c}{m_t} \right |_{M_X}  \; , \nonumber \\
\left |\hat{V}_{ij} \right |_{M_Z} & = & \frac{1}{\xi_t ~ \xi_b} ~ 
\left |\hat{V}_{ij} \right |_{M_X} \; 
\end{eqnarray}
with $(ij) = (ub)$, $(cb)$, $(td)$ or $(ts)$. In the same approximations, 
the renormalization group equations for the Yukawa coupling eigenvalues 
$f_t$, $f_b$ and $f_{\tau}$ read \cite{Babu}:
\begin{eqnarray}
16 \pi^2 \frac{{\rm d} f_t}{{\rm d} \chi} & = & f_t \left ( 6 f^2_t ~ + ~ 
f^2_b ~ - ~ G_{\rm u} \right ) \; , \nonumber \\
16 \pi^2 \frac{{\rm d} f_b}{{\rm d} \chi} & = & f_b \left ( f^2_t ~ + ~ 
6 f^2_b ~ + ~ f^2_{\tau} ~ - ~ G_{\rm d} \right ) \; , \nonumber \\
16 \pi^2 \frac{{\rm d} f_{\tau}}{{\rm d} \chi} & = & f_{\tau} \left (
3 f^2_b ~ + ~ 4 f^2_{\tau} ~ -~ G_{\rm e} \right ) \; ,
\end{eqnarray}
where the quantities $G_{\rm n}$ have been given in Eq. (4.2).

\vspace{0.3cm}

With the typical inputs $m_t (M_Z) \approx 180$ GeV, $m_b (M_Z) \approx 3.1$ GeV, 
$m_{\tau} (M_Z) \approx 1.78$ GeV and those listed in Table 1,
we calculate $\xi_t$ and $\xi_b$ for arbitrary $\tan\beta_{\rm susy}$
by use of the above equations. Our result is illustrated in Fig. 4.
We see that $\xi_b \approx 1$ for $\tan\beta_{\rm susy} \leq 10$. This
justifies our approximation made previously in deriving 
$\hat{M}^{\rm u}_0 (M_Z)$ and $\hat{M}^{\rm d}_0 (M_Z)$. 
Within the perturbatively allowed region of 
$\tan\beta_{\rm susy}$ \cite{Froggatt},
$\xi_b$ may be comparable in magnitude with $\xi_t$ when $\tan\beta_{\rm susy}
\geq 50$. In this case, the evolution effects of quark mass matrices and
flavor mixings are sensitive to both $f_t$ and $f_b$. 

\vspace{0.3cm}

Clearly the analytical results of $|\hat{V}_{us}|$, $|\hat{V}_{cd}|$,
$|\hat{V}_{ub}/\hat{V}_{cb}|$ and $|\hat{V}_{td}/\hat{V}_{ts}|$ 
as those given in Eqs. (3.7) and (3.9) are almost scale-independent,
i.e., they hold at both $\mu=M_X$ and $\mu=M_Z$. Non-negligible
running effects can only appear in the expression of $|\hat{V}_{cb}|$ or 
$|\hat{V}_{ts}|$, which is a function of the mass ratios $m_s/m_b$ and 
$m_c/m_t$ (see Eq. (3.8) for illustration). With the help of Eq. (4.9),
we find the renormalized relation between $|\hat{V}_{cb}|$ (or $|\hat{V}_{ts}|$)
and the quark mass ratios at the weak scale $M_Z$:  
\begin{equation}
|\hat{V}_{cb}| \; \approx \; |\hat{V}_{ts}| \; \approx \;
\sqrt{2} \left ( \xi^2_b \frac{m_s}{m_b} - \xi^2_t \frac{m_c}{m_t} \right )
\left [ 1 + 3 \xi_t \xi_b \left ( \xi^2_b \frac{m_s}{m_b} + \xi^2_t \frac{m_c}{m_t}
\right ) \right ] \; .
\end{equation}
This result will recover that in Eq. (3.8) if one takes $M_Z=M_X$ (i.e.,
$\xi_t=\xi_b=1$). Using $m_b/m_s=34\pm 4$ \cite{Narison}
and taking $m_c/m_t=0.005$ typically,
we confront Eq. (4.11) with the experimental data on $\hat{V}_{cb}$
(i.e., $|\hat{V}_{cb}|=0.0388\pm 0.0032$ \cite{Neubert}). 
As shown in Fig. 5, our result is in good agreement with experiments 
for $\tan\beta_{\rm susy} < 50$. This implies that
the quark mass pattern $\hat{M}_0$ or $\hat{M}_{\rm H}$,
proposed at the supersymmetric GUT scale $M_X$, may have a large chance to 
survive for reasonable values of $\tan\beta_{\rm susy}$.

\vspace{0.3cm}

Note that evolution of the $CP$-violating parameter $J$ is dominated by 
that of $|\hat{V}_{cb}|^2$. Note also that
three sides of the unitarity triangle $\hat{V}^*_{ub}\hat{V}_{ud} +
\hat{V}^*_{cb}\hat{V}_{cd} + \hat{V}^*_{tb}\hat{V}_{td}=0$ have 
identical running effects from $M_X$ to $M_Z$, hence its three
inner angles are scale-independent and take the same values as those
given in Eq. (3.10) or Fig. 3. As a result, measurements of $\alpha$,
$\beta$ and $\gamma$ in the forthcoming experiments of $B$ physics
may check both the quark mass {\it Ansatz} at the
weak scale and that at the supersymmetric GUT scale.

\section{Summary}
\setcounter{equation}{0}

Without the assumption of specific mass matrices, we have pointed out
that part of the observed properties of flavor mixings can be well understood
just from the quark mass hierarchy. In the quark mass limits such as
$m_u=m_d=0$, $m_t\rightarrow \infty$ or $m_b\rightarrow \infty$, 
a few instructive relations between the KM matrix
elements and quark mass ratios are suggestible from current experimental data.
We stress that such {\it Ansatz}-independent results may serve as 
a useful guide in constructing the specific quark mass matrices at either 
low-energy scales or superheavy scales. 

\vspace{0.3cm}
 
Starting from the flavor permutation symmetry and assuming an explicit pattern
of symmetry breaking, we have proposed a new quark mass {\it Ansatz} 
at the weak scale. We find that all experimental
data on quark mixings and $CP$ violation can be accounted for 
by our {\it Ansatz}. In particular, we obtain an
instructive relation among $|V_{cb}|$, $m_s/m_b$ and $m_c/m_t$ in the
next-to-leading approximation (see Eq. (3.8)). The scale-independent
predictions of our quark mass pattern, such as 
$0.18 \leq \sin(2\alpha) \leq 0.58$, $0.5\leq \sin(2\beta) \leq 0.78$
and $-0.08 \leq \sin(2\gamma) \leq 0.5$, can be confronted with
the forthcoming experiments at KEK and SLAC $B$-meson factories.

\vspace{0.3cm}

With the same {\it Ansatz} prescribed at the supersymmetric GUT scale
$M_X$, we have derived the renormalized quark mass matrices at the weak
scale $M_Z$ for small $\tan\beta_{\rm susy}$ and calculated the renormalized 
flavor mixing matrix elements at $M_Z$ for arbitrary $\tan\beta_{\rm susy}$.
Except $|\hat{V}_{cb}|$ and $|\hat{V}_{ts}|$, the other asymptotic
relations between the KM matrix elements and quark mass ratios
are almost scale-independent. We find that the renormalized 
result of $|\hat{V}_{cb}|$ (or $|\hat{V}_{ts}|$) is 
in good agreement with the relevant experimental data for reasonable 
values of $\tan\beta_{\rm susy}$.

\vspace{0.3cm}

In this work we neither assumed a specific form for the charged lepton 
mass matrix nor supposed its relation with the down quark mass matrix within the
supersymmetric GUT framework. Of course, this can be done by following
the strategy proposed in Ref. \cite{Georgi}. Then one may obtain the
relations between $m_d$, $m_s$, $m_b$ and $m_e$, $m_{\mu}$, $m_{\tau}$.
Such an {\it Ansatz}, based on the specific GUT scheme and flavor 
permutation symmetry breaking, will be discussed somewhere else.

\vspace{0.5cm}

\begin{flushleft}
{\Large\bf Acknowledgements}
\end{flushleft}

The author would like to thank A.I. Sanda for his warm hospitality and the Japan
Society for the Promotion of Science for its financial support. He is also
grateful to H. Fritzsch, A.I. Sanda and K. Yamawaki for their useful comments
on the topic of permutation symmetry breaking and on part of this work.

\end{document}